\documentclass[12pt]{article}
\usepackage{graphicx}
\begin{document}
Birth, survival and death of languages by Monte Carlo simulation

\bigskip

\centerline{C. Schulze$^1$, D. Stauffer$^1$, S. Wichmann$^2$}

\bigskip
\centerline{$^1$ Institute for Theoretical Physics, Cologne University, D-50923 K\"oln, 
Euroland}

\bigskip
\centerline{$^2$ Department of Linguistics, Max Planck Institute for 
Evolutionary Anthropology}
\centerline{Deutscher Platz 6, D-04103 Leipzig, Germany}

\bigskip
Abstract: Simulations of physicists for the competition between adult languages
since 2003 are reviewed. How many languages are spoken by how many people? How many 
languages are contained in various language families? How do language 
similarities decay with geographical distance, and what effects do natural
boundaries have? New simulations of bilinguality are given in an appendix.

\bigskip
\section{Introduction}

While the emergence and learning of human languages has been simulated since 
decades on computers \cite{www}, and while a later economics Nobel laureate 
also contributed to linguistics long ago \cite{selten}, the competition 
between existing languages of adults is a more recent research trend, where 
physicists have tried to play a major role. It follows the principle of survival
of the fittest, as known from Darwinian evolution in biology, and indeed many
of the techniques have been borrowed from simulational biology \cite{newbook}.
This emphasis from physics on the competition of existing languages for adult 
humans started with Abrams and Strogatz\cite{abrams} and was then followed 
by at least six groups independently 
\cite{spain,kosmidis,schulze,argentina,schwammle,viviane}.
More recently, of course, reviews\cite{reviews,newbook} and conferences
brought them together, and others followed them \cite{tibi,mallorca,tuncay}. 

Today about 7000 different languages (as defined by linguists) are spoken,
and about every ten days one of them dies out. On the other hand, the split
of Latin into different languages spoken from Portugal to Romania is well
documented. In statistical physics, we can describe and explain the 
pressure which air molecules of a known density and temperature exert on the 
walls. But we cannot predict were one given molecule will be one second from 
now. Similarly, the application of statistical physics tools to linguistics 
may describe the ensemble of the seven thousand presently existing languages, 
but not the extinction of one given language in one given region on Earth. 
Figure 1 shows how many languages exist today, as a function of the number of 
speakers of that language. A statistical theory of language competition thus 
first of all should try to reproduce such results, in order to validate the 
model. If if fails to describe this fact, why should one trust it at all?
Or as stated by linguist Yang on page 216 of \cite{yang}: It is time for the 
ancient field of linguistics to join the quantitative world of modern science.

This review starts with our own model for numerous languages in section 2,
followed by a review of the alternative model of Viviane de Oliveira and 
coworkers
\cite{viviane}. Then we review more shortly the many other models which at 
present do not allow the simulation of thousands of different languages. 

\begin{figure}
\begin{center}
\includegraphics[angle=-90,scale=0.5]{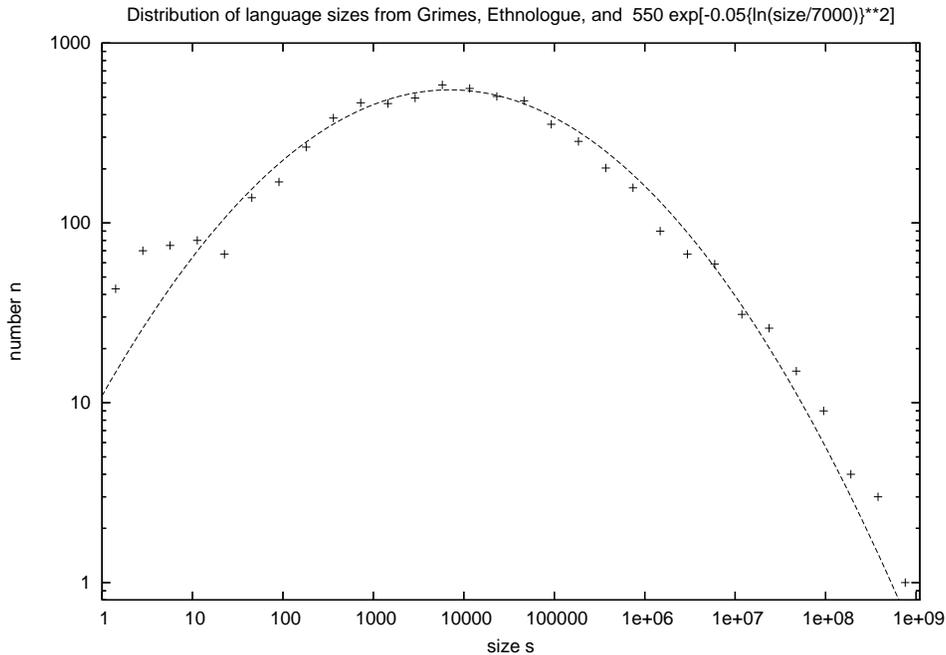}
\end{center}
\caption{
Empirical variation of the number $N_s$ of languages spoken by $s$ people 
each. For better presentation, the language sizes $s$ are binned in powers of
two. Data from Ethnologue \cite{ethnologue}, as plotted in \cite{langsol}.
The parabola corresponds to a log-normal distribution; we see deviations 
from  it for the smallest sizes \cite{sutherland}.
}
\end{figure}

\section{Schulze Model}
\subsection{Definition}

Our own simulations, also called the Schulze model, characterise each 
language (or grammar) by $F$ independent features each of which can 
take one of $Q$ different values; the binary case $Q=2$ allows the storage
in bit-strings. Three basic mechanisms connected with probabilities $p, \;
q$ and $r$ are common to all variants:
\bigskip

i) With probability $p$ at each iteration and for each feature, this 
feature is changed (or mutated in biological language). This change is 
random or not, depending on process ii).

ii) With probability $q$ the mutation/change under i) is not random but 
instead transfers the value of this feature from another person in the
population. This transfer is called diffusion by linguists. With probability
$1-q$, the change is random.

iii) With probability $(1-x)^2r$ (also $(1-x^2)r$ has been used instead) 
somebody discards the
mother language and takes over the whole language (all $F$ features)
from another person in the population. Here $x$ is the fraction of people
speaking the old language. This flight is called shift by linguists.
\bigskip

Several variants are possible: One can use one joint population where everybody
can meet everybody for transfer and shift; or we put people onto a square $L 
\times L$ lattice or more complicated network, where diffusion and/or shift 
are possible only from a randomly selected neighbour. People may migrate
on this lattice, which physicists would call diffusion. The population
can be fixed, meaning that at every iteration all adults are replaced by
their children. Or it can grow by a suitable birth and death process;
in this case the shifting probability can include also a factor 
proportional to the population. If one dislikes to have three free 
parameters $p,q,r$ one may set $q=r=1$ without much loss in results.

For the number $F$ of features, from 8 to 64 were used in simulations.  Real 
languages contain many thousands of words for everyday use, and thus one should 
identify one feature rather with an independent grammatical element (like
the order of subject, object and verb in a sentence) than with a word.
$F$ for real languages was estimated as about 30 \cite{briscoe} or about 40 
to 50 \cite{yang} such choices, and  the Word Atlas of Language Structures
\cite{wals} lists 138 features with up to $Q = 9$ values. These grammar
sizes thus correspond roughly to what has been simulated.
According to \cite{wichholman} the rate of change in normal
linguistic typological features, i.e., excluding a few extraordinarily
unstable ones, is 16 \% per 1000 years.

% From inspecting average grammatical differences among languages (in 
%terms of the features represented \cite{wals}) at different 
%degrees of taxonomic distances it appears to be the case that 
%differences within languages, i.e., among dialects, are on average 20 \%. 
%Assuming that it takes a language 1000 years to develop, we might expect 
%1 \% difference per 50 years, i.e., roughly a 1 \% change per human generation.
%This is admittedly a crude estimate, but lends a degree of realism to 
%our simulations.

\subsection{Results}
If we start with everybody speaking one language (or with just one Eve), 
then at low $p$ this language still dominates and is spoken by more than 
half of the total population, with the remaining people speaking a minor and 
short-lived variant of this dominating language. At high $p$, on the other
hand, the whole population soon fragments into many languages, roughly such 
that everybody selects nearly randomly one of the $Q^F$ possible languages.
This corresponds to the biblical story of the Tower of Babel. We thus have
dominance for small $p$ and fragmentation for large $p$, with a first-order
phase transition or jump at some threshold value which depends on the other
parameters and details of the model.

If instead we start with everybody speaking a randomly selected language,
then for high $p$ this situation remains. For low $p$, however, after some 
time one language by random accidents happens to grow to a sufficiently large 
size such that it then grows rapidly to be spoken by more than half of the 
population. Thus a transition from initial fragmentation to final dominance
happens in Fig.2. The threshold value is different from the one for the 
opposite direction from  dominance to fragmentation: we have hysteresis as is 
common for first-order transitions, Fig.3. Empirically, this transition to one 
dominating language was observed on the American continent; within the last
five centuries, two thirds of the native Brazilian languages have died out. 
And in the last half century we observed the rise of English in physics research
publications. While 85 years ago, physicist Bose sent from India his paper
to Einstein in order to have it translated from English to German (which
lead to Bose-Einstein condensation), after World
War II physics research was usually published in English, first in Japan, since
the 1960's in (West) Germany, a decade later in France, since the 1990's in 
Russia; finally, China has witnessed a surge in physics papers written in 
English since 2000.

\begin{figure}
\begin{center}
\includegraphics[angle=-90,scale=0.38]{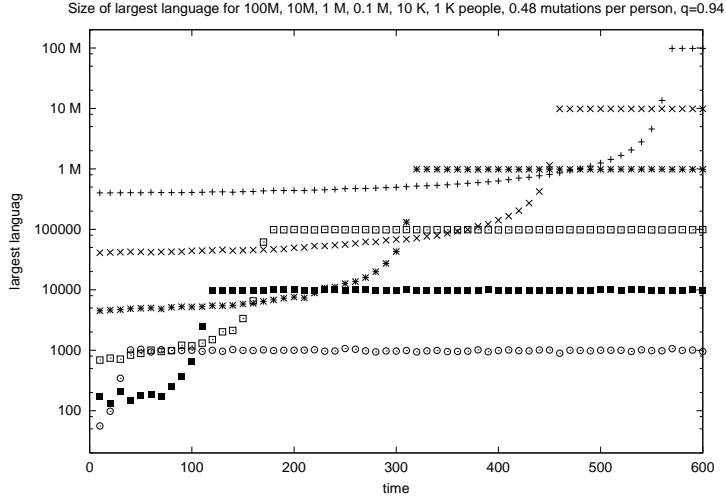}
\end{center}
\caption{
Variation with time of the number of people speaking the most widespread 
language, for various population sizes. The larger the population is, the 
longer is the time until the transition from fragmentation to dominance takes 
place. $Q=2, F=8, p = 0.06, q = 0.94;$ from \cite{reviews}.
}
\end{figure}

\begin{figure}
\begin{center}
\includegraphics[angle=-90,scale=0.38]{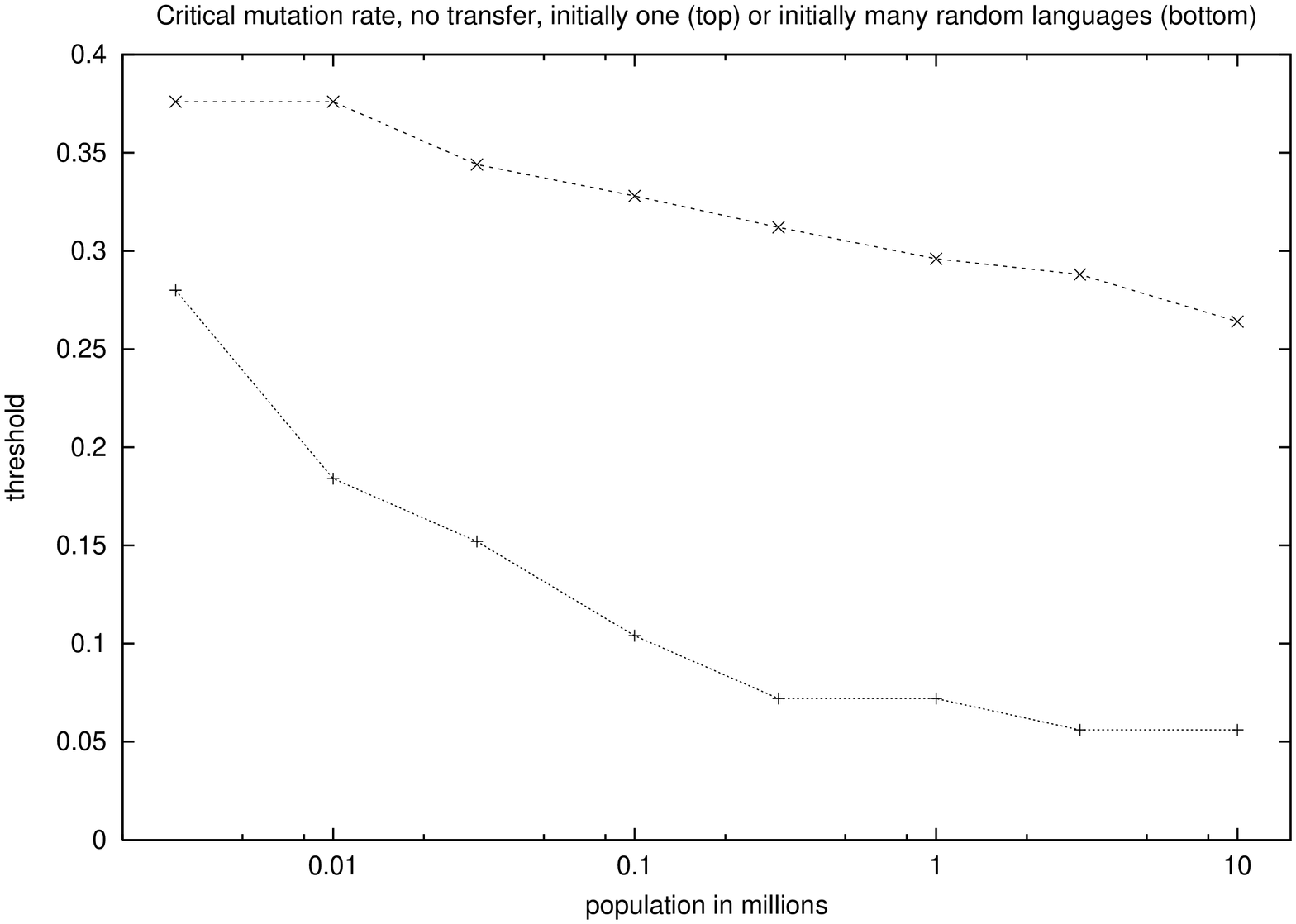}
\end{center}
\caption{
Dependence of the mutation threshold for the phase transition on the population 
size; upper data from dominance to fragmentation, lower data
from fragmentation to dominance. Above the curves we arrive at fragmentation,
below at dominance. From \cite{reviews}.
}
\end{figure}

The time needed to go from fragmentation to dominance increases roughly 
logarithmically with population size, at least in the binary case $Q=2$ without 
lattice. Thus a mathematical solution for an infinite population might never 
get this transition. In other words, proper models should be agent-based 
\cite{fent}, with individuals acting on their own; one should not average over
the whole population, using differential equation for the concentrations.
Such simulations have been standard in computational physics for half a century
(Monte Carlo and Molecular Dynamics), while mean field approximations
average over many individuals and can give somewhat or completely wrong 
results. (The transition from fragmentation to dominance may require a 
shift probability $(1-x)^2r$ instead of $(1-x^2)r$.)

The language size distribution to be compared with Fig.1 is shown in Fig.4a. To
get it, we looked at non-equilibrium results and introduced random 
multiplicative noise, since otherwise the language sizes were too small and 
their distribution too irregular. Fig.4b avoids these tricks and instead
places the people on a directed scale-free network, discussed below.

\begin{figure}
\begin{center}
\includegraphics[angle=-90,scale=0.35]{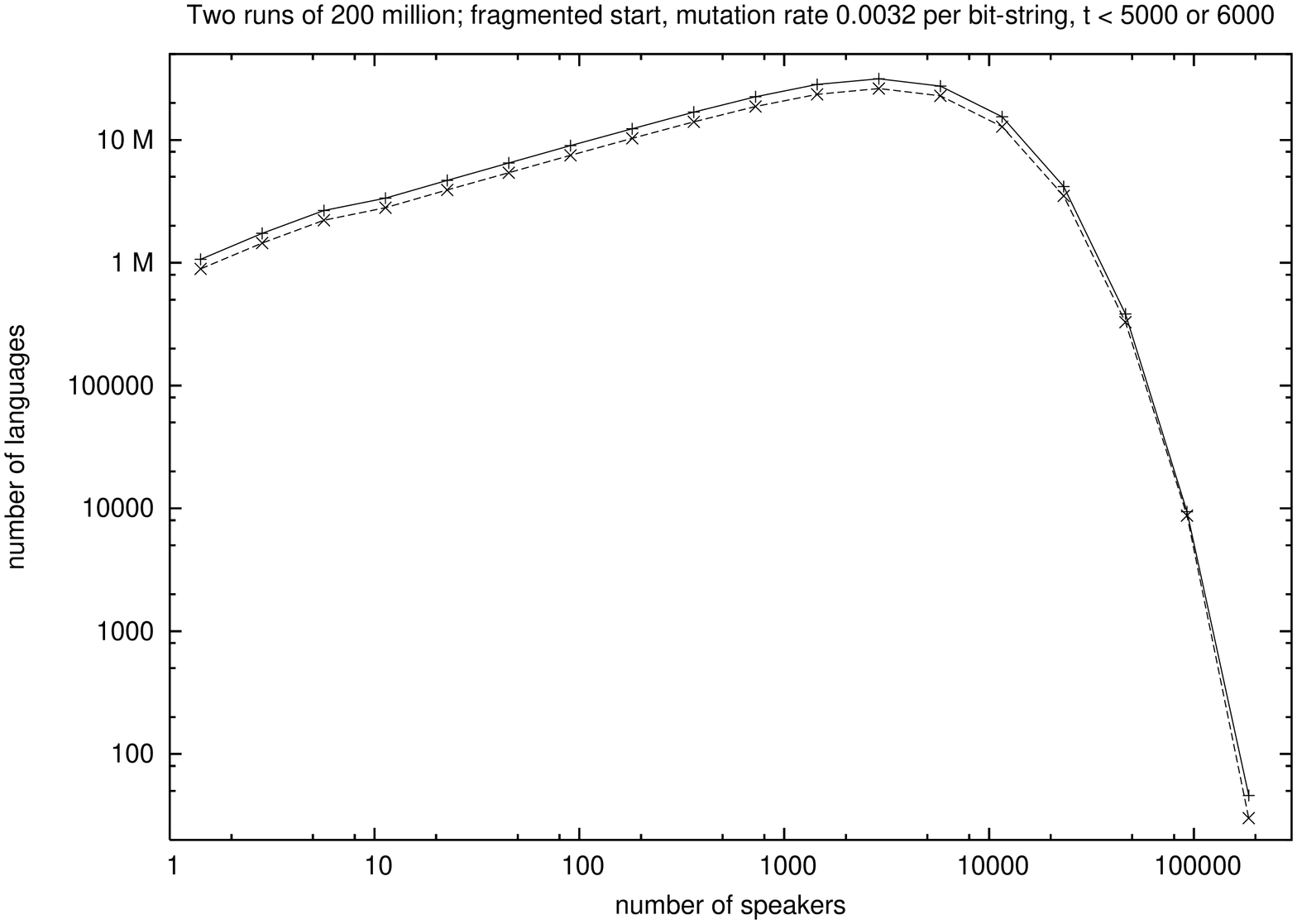}
\includegraphics[angle=-90,scale=0.35]{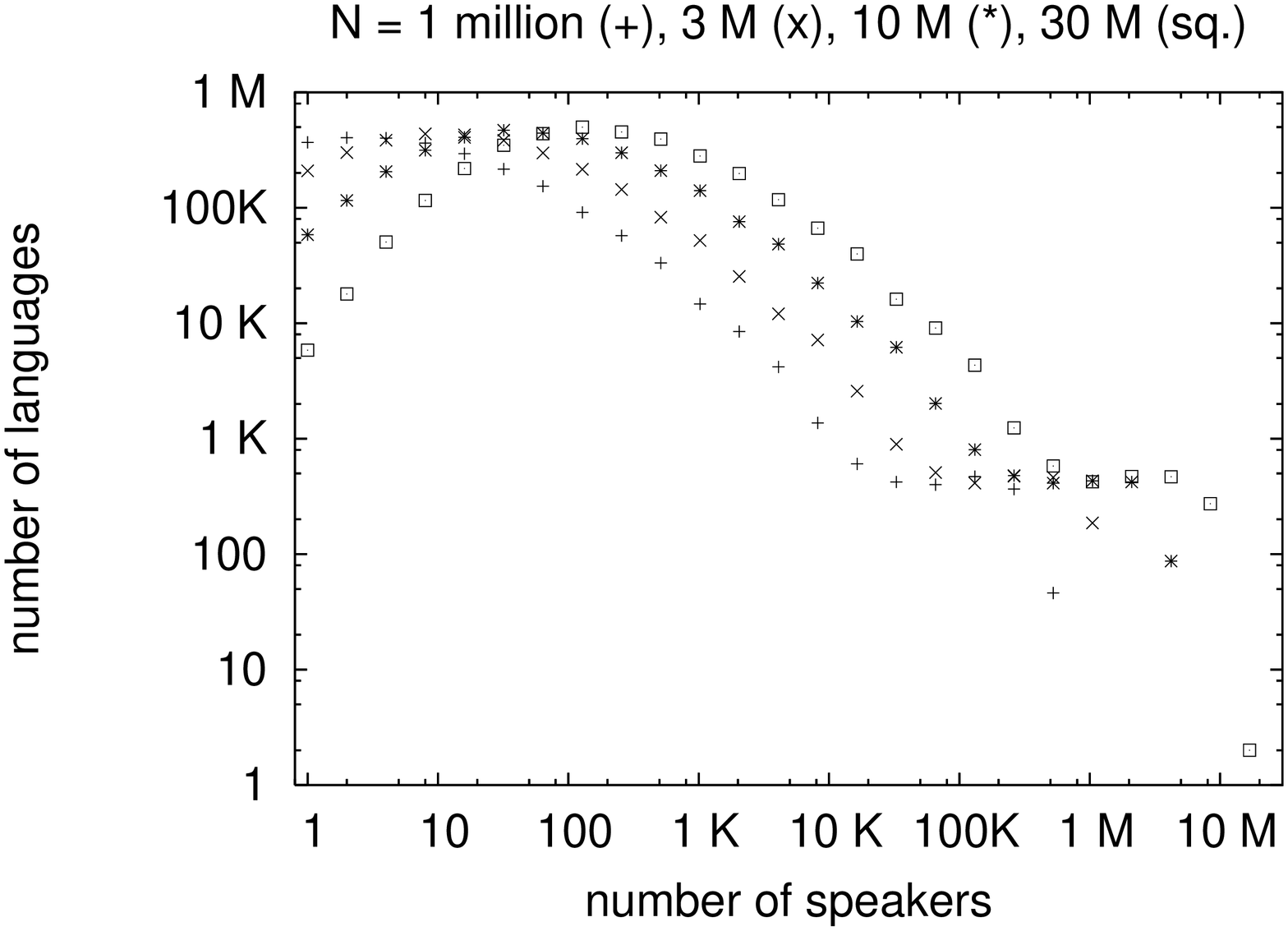}
\end{center}
\caption{
Language size distribution $N_s$. Top: without lattice, with random 
multiplicative noise, not in equilibrium \cite{langsol}. Bottom: 
On scale-free network in equilibrium \cite{patt}.}

\end{figure}

No lattice or other spatial structure was employed in the above simulations. On 
a lattice one can look at language geography \cite{goebl,holman,cavalli}. North
and South of the Alps, different languages are spoken, and the same separation
is made by the English Channel. Genetic and linguistic boundaries in 
Europe mostly coincide, and about two thirds of them agree with natural
boundaries like a mountain chain or sea \cite{sokal}. We simulated this effect
on a lattice \cite{barrier} with contact only between nearest neighbours and
a horizontal barrier separating the upper from the lower half. The
shift from a small to a large language happens across the barrier only with a 
small crossing probability $c$. For $c=0$ one thus has two completely separated
halves of the lattice, and trivially the languages which evolve as dominating
are different on both sides of the border. With $c=1$ the border has no effect,
and only one language dominates. Fig.5 shows how often for small but finite
$c$ two separate dominating languages may coexist; already quite small $c$
suffice, particularly for large lattices, to unify the two regions into only 
one with the same language dominating on both sides.

\begin{figure}
\begin{center}
\includegraphics[angle=-90,scale=0.40]{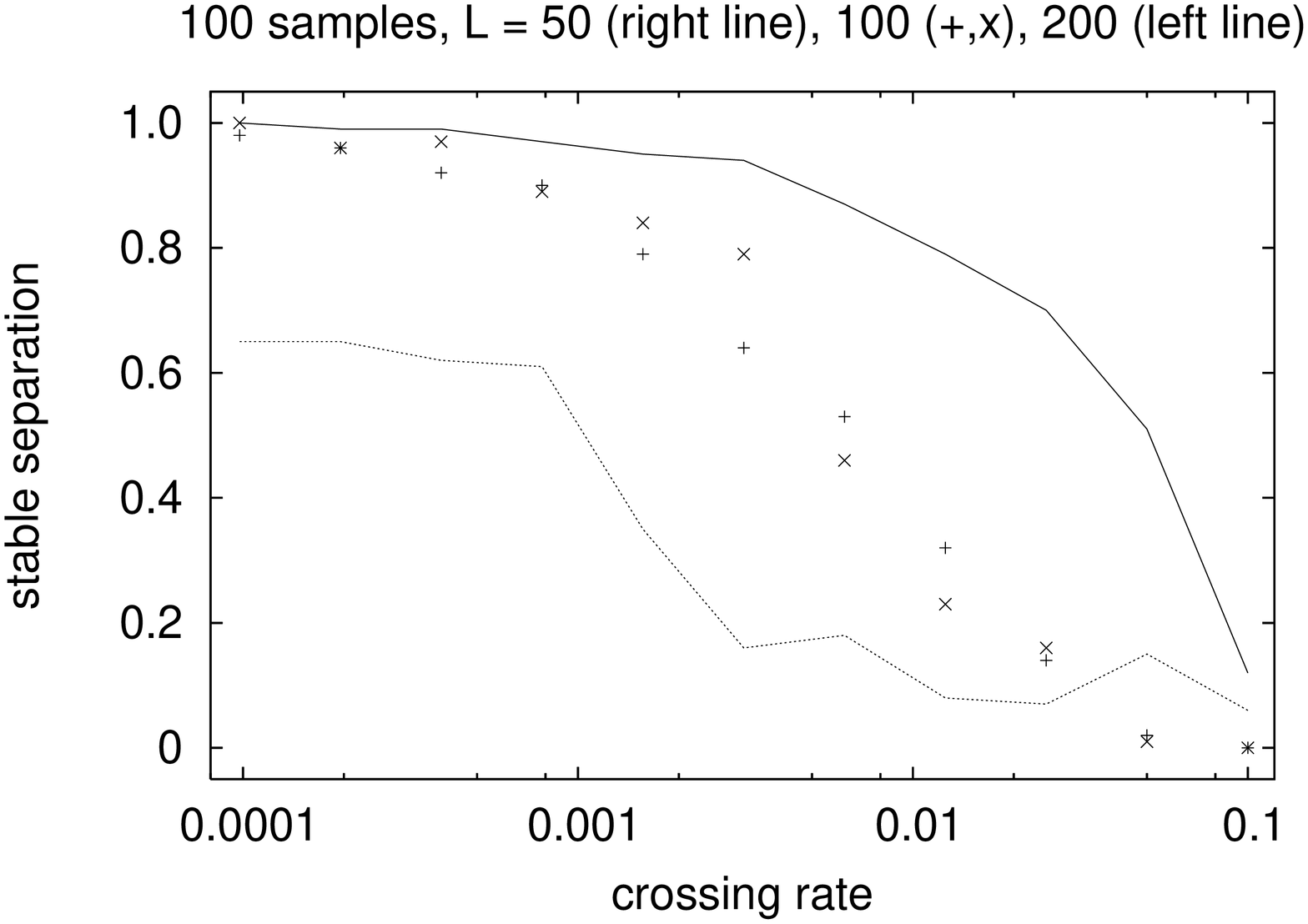}
\end{center}
\caption{
Fraction of cases when a semi-permeable barrier allowed two different languages
to dominate on its two sides in the Schulze model.
}
\end{figure}

Fig.4b employed a directed Barab\'asi-Albert scale-free 
network. These networks are grown from a small fully connected core such that 
each new network member selects $m$ already existing members as teachers.
The more people have selected a certain teacher before, the higher is the
probability that this teacher will again be selected. Information only flows 
from the teacher to the person who selected this teacher, not in the opposite 
direction \cite{ba}. 
\section{Viviane Model}
\subsection{Definition}

The model of Viviane de Oliveira et al \cite{viviane} has become known as the 
Viviane model (following the Brazilian tradition of how to call people).
It simulates the colonisation of an uninhabited continent by people.
Each site $j$ of an $L \times L$ square lattice can later be populated
by $c_j$ people; this carrying capacity $c_j$ is selected randomly between
1 and some $m \sim 10^2$. Initially only one site $i$ is occupied by
$c_i$ people. 

Then at each time step, one randomly selected empty lattice neighbour $j$ of
occupied sites becomes occupied with probability $c_j/m$ by $c_j$ people. Thus
after some time the whole lattice becomes occupied and the simulation stops.
In contrast to the Schulze model, the Viviane model is a growth process and
not one eventually fluctuating about some equilibrium.

Languages have no internal structure and are simply numbered 1, 2, 3,,,
with 1 being the number of the language spoken on the originally occupied site.
All people within one lattice site speak the same language. First, if a new 
site has been colonized the language spoken there is taken from one
of the occupied neighbours $k$, proportional to the fitness $F_k$ of that 
neighbour site $k$. This fitness is the total number of people speaking
anywhere in the lattice the language of $k$, except that it is bounded 
from above by a maximum fitness $M_k$ fixed randomly between 1 and some 
$M_{\max} \sim 10^3$.  Second, mutations (language change) are made with 
probability $\alpha/F_j$ on the freshly occupied site $j$ only, from the 
selected language of neighbour $k$.  A mutation means that a new language 
is created which gets a new number not used previously.

In this way, the flight (shift) from small language and the mutations,
which were two separate processes in the Schulze model, are combined into
one process; and this process also is a transfer (diffusion) process which
in the Schulze model was dealt with separately. Thus we have here only
one free parameter $\alpha$, the mutation coefficient, instead of three
parameters $p,q,r$ in the Schulze model. 

Variants allow mutations also later, after a site is occupied. Or a
language is characterised by a string of $F$ bits ($Q=2$ in the Schulze 
notation) and only different bit-strings count as different languages
\cite{pmco}. Or the capacities $c_k$ are not homogeneously distributed
between 1 and $m$ but more often small than large, with a frequency
proportional to $1/c_j$, as long as it is not larger than the maximum $m$.
Lot of computer time then is saved if after the occupation of the new sites 
one selects two of its occupied neighbours and takes the language from the 
one with the bigger capacity; if only one neighbour is occupied then its
language is taken over.

\subsection{Results}

\begin{figure}
\begin{center}
\includegraphics[angle=-90,scale=0.4]{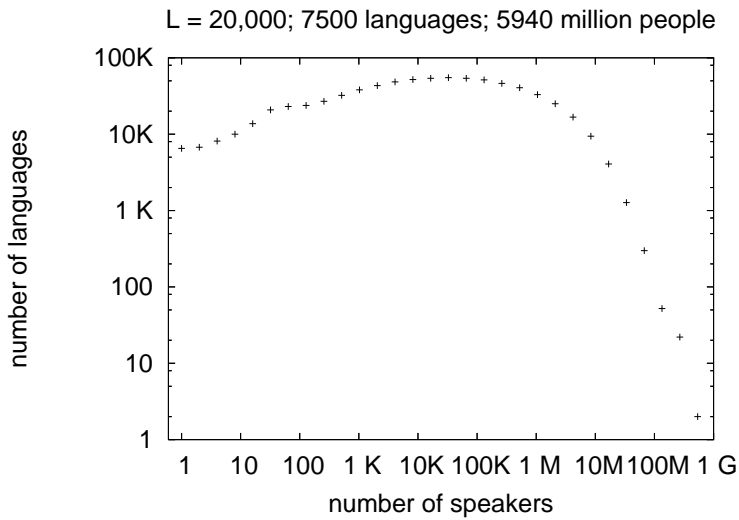}
\end{center}
\caption{
Variation of the number $N_s$ of languages spoken by $s$ people 
each in the Viviane model, as modified and published in \cite{pmco}. 
}
\end{figure}

In contrast to the Schulze model, the Viviane model gives languages spoken by
$10^9$ people if a sufficiently large lattice is used. The language size 
distribution $N_s$ has a maximum at moderately small language sizes $s$. 
However, instead of a round parabola as in Fig.1, the log-log plot of $N_s$ 
versus $s$ gives two straight lines meeting at the maximum, that means
one power law for small $s$ (where $N_s$ increases with $s$) and another
power law for large $s$ (where $N_s$ decreases with $s$). So, not everything
is solved yet.

Crucial progress was made by Paulo Murilo de Oliveira (not the same family as
Viviane de Oliveira), who introduced the above-mentioned  modifications \cite{pmco}: 
Languages (grammars) are characterized by bit-strings ($Q=2$) of length $F 
\simeq 13$ and count as different only when their bit-strings differ;
the carrying capacities $c$ are selected with a probability $\propto 1/c$, 
and the newly colonized site gets the fitter language of two previously
occupied neighbours. Now the distribution is roughly log-normal, Fig.6, with 
enhancement for very small sizes; the total population and the total number 
of languages can be made close to the present reality, the maximum
of the parabola in a double-logarithmic plot (with binning by factors of
two in $s$) is near $s \simeq 10^4$, while the largest language is spoken
by $10^9$ people, similar to Mandarin Chinese. 

Fig.7 shows for both the modified Viviane model and the Schulze lattice model 
that languages in general are less similar to each other if they are widely 
separated geographically, in agreement with reality \cite{holman,cavalli}.
Note the difference in scales: One lattice constant (distance between nearest 
neighbours) corresponds to about one kilometer in the Viviane model and 1000
kilometers in the Schulze model, if Fig.7 is compared to reality \cite{holman}.

Also the classification of different languages into one family, like the 
Indo-European languages, has been simulated with moderate success. Following 
the history of the mutations during the colonisation, a language tree like
can be constructed in the unmodified version (Fig.11.15 in \cite{reviews}). 
One can imagine that this is Latin, splitting up into Romanian, Italian, 
Spanish and French, with Spanish then splitting into Castellan, Galego and 
Catalan, and Catalan mutating further into Mallorquin. (Many small branches 
were omitted for clarity.) More quantitative information is 
obtained from the modified bit-string version \cite{pmco}.  The mutated language
on a newly occupied site starts a new family if it differs in two or more bits 
from the bit-string characterising the historically first language 
of the old family. The size distribution of language families in Fig.8 agrees
in its central part with the empirically observed \cite{wichmann} exponent 
--0.525 and is independent of the length $F = 8$, 16, 32 or 64 of the 
bit-strings for the Viviane model and independent of the population size for
the Schulze model.

\begin{figure}
\begin{center}
\includegraphics[angle=-90,scale=0.28]{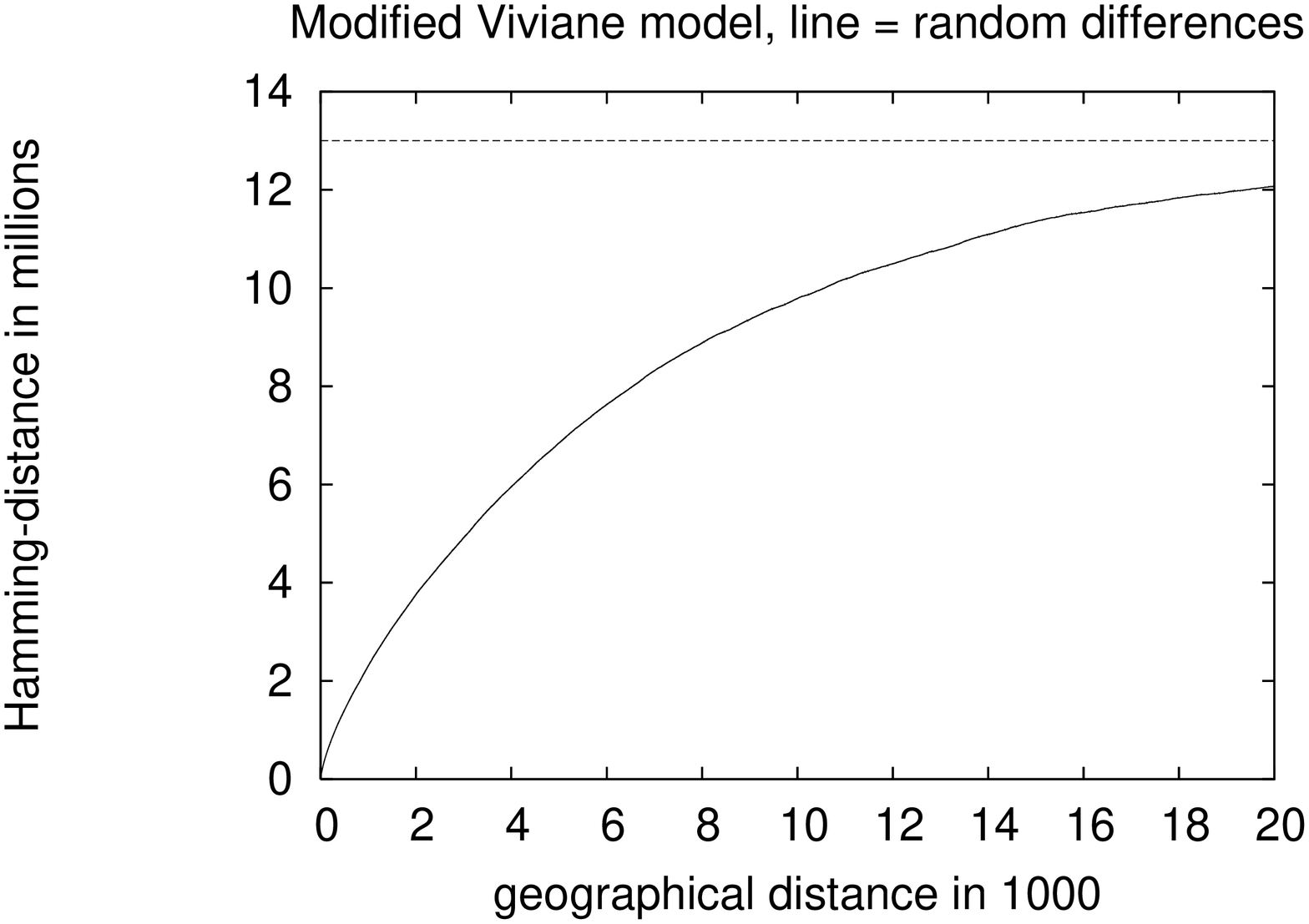}
\includegraphics[angle=-90,scale=0.28]{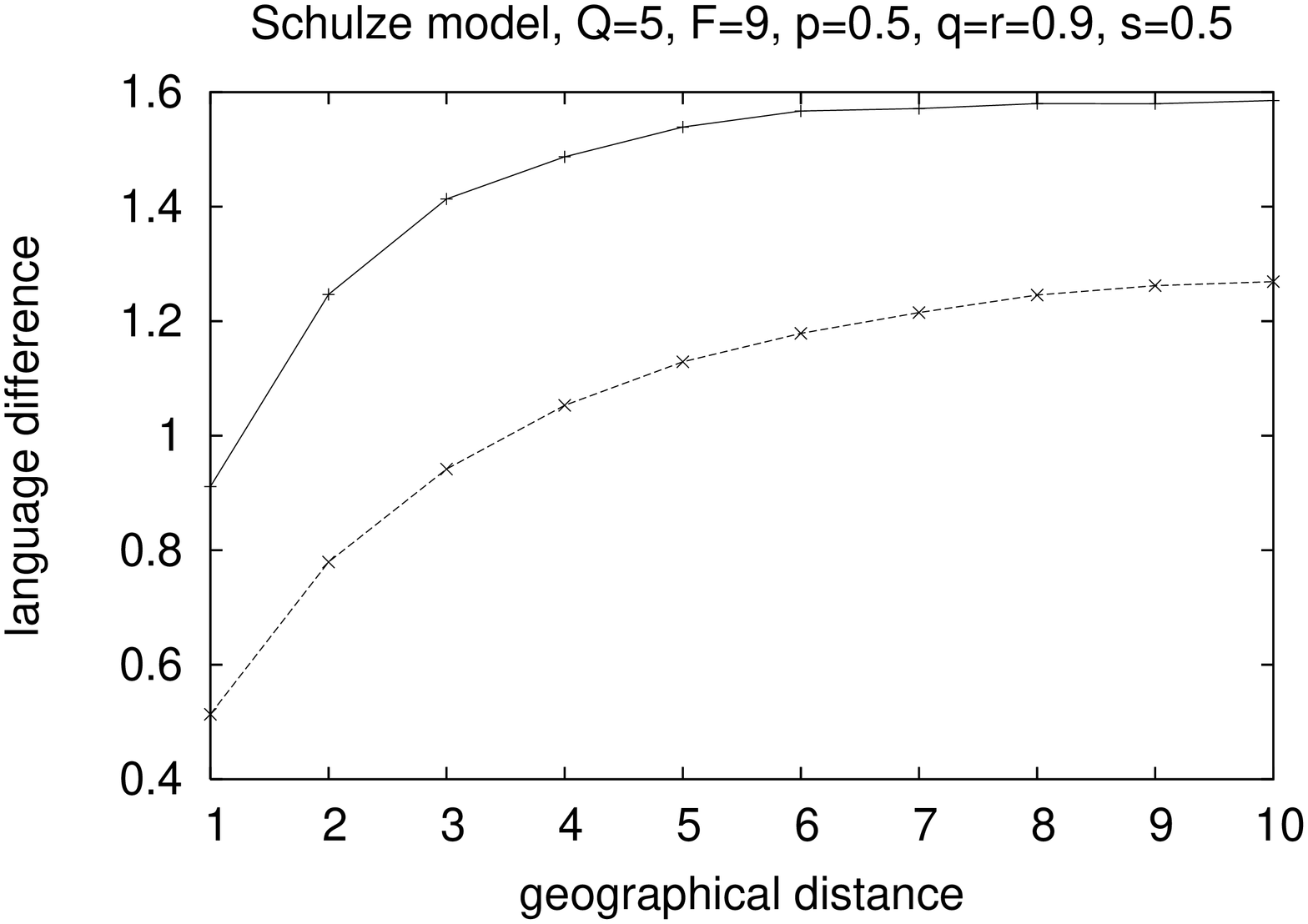}
\end{center}
\caption{
Language differences (in arbitrary units \cite{holman}) as a function
of geographical distance in the Viviane model (top) and the Schulze model
(bottom). The horizontal line corresponds to completely uncorrelated
languages. In the bottom part, + and x correspond to start with fragmentation
(+) and with dominance (x). From \cite{holman,patt}.
}
\end{figure}

\begin{figure}
\begin{center}
\includegraphics[angle=-90,scale=0.35]{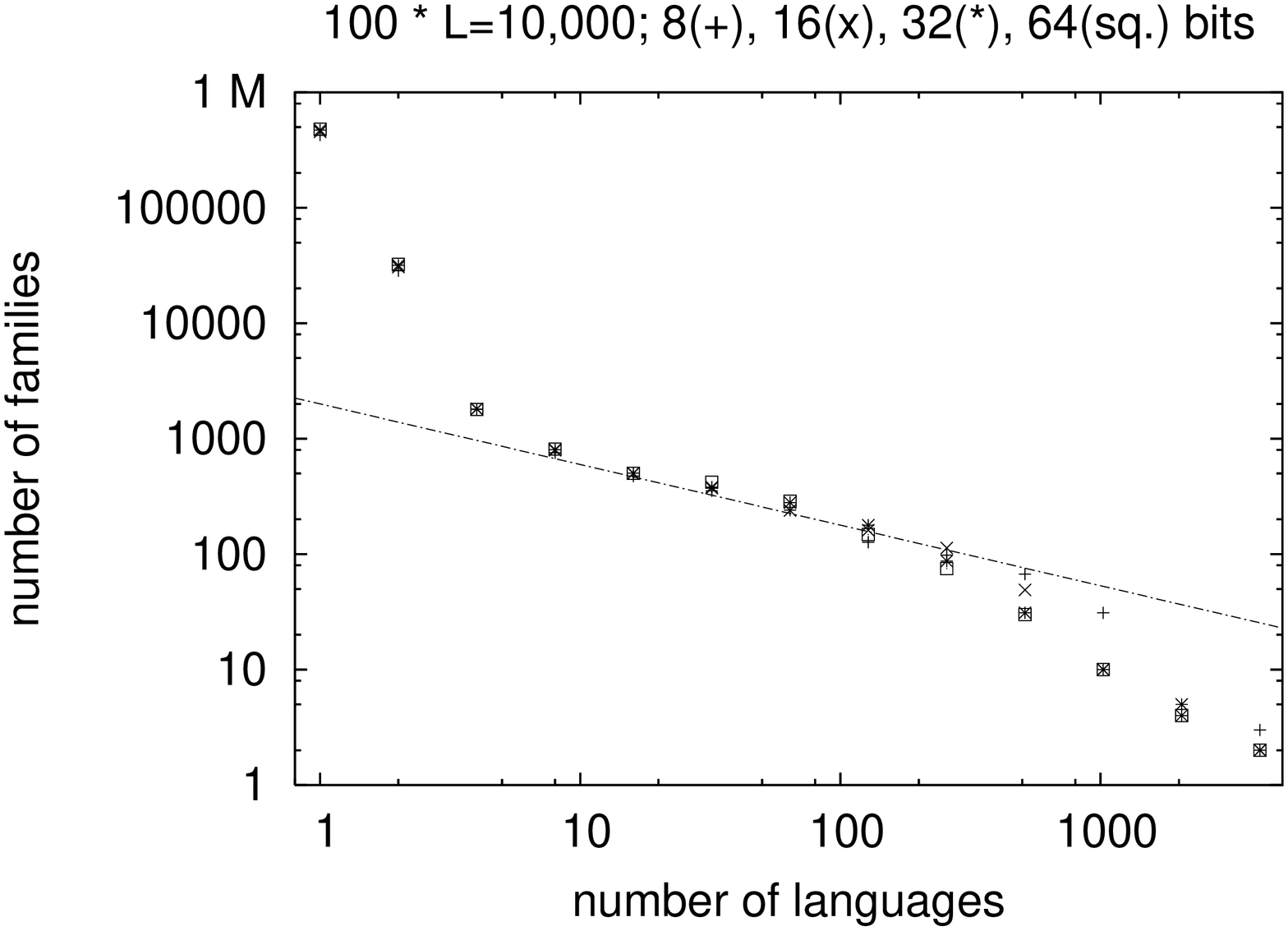}
\includegraphics[angle=-90,scale=0.35]{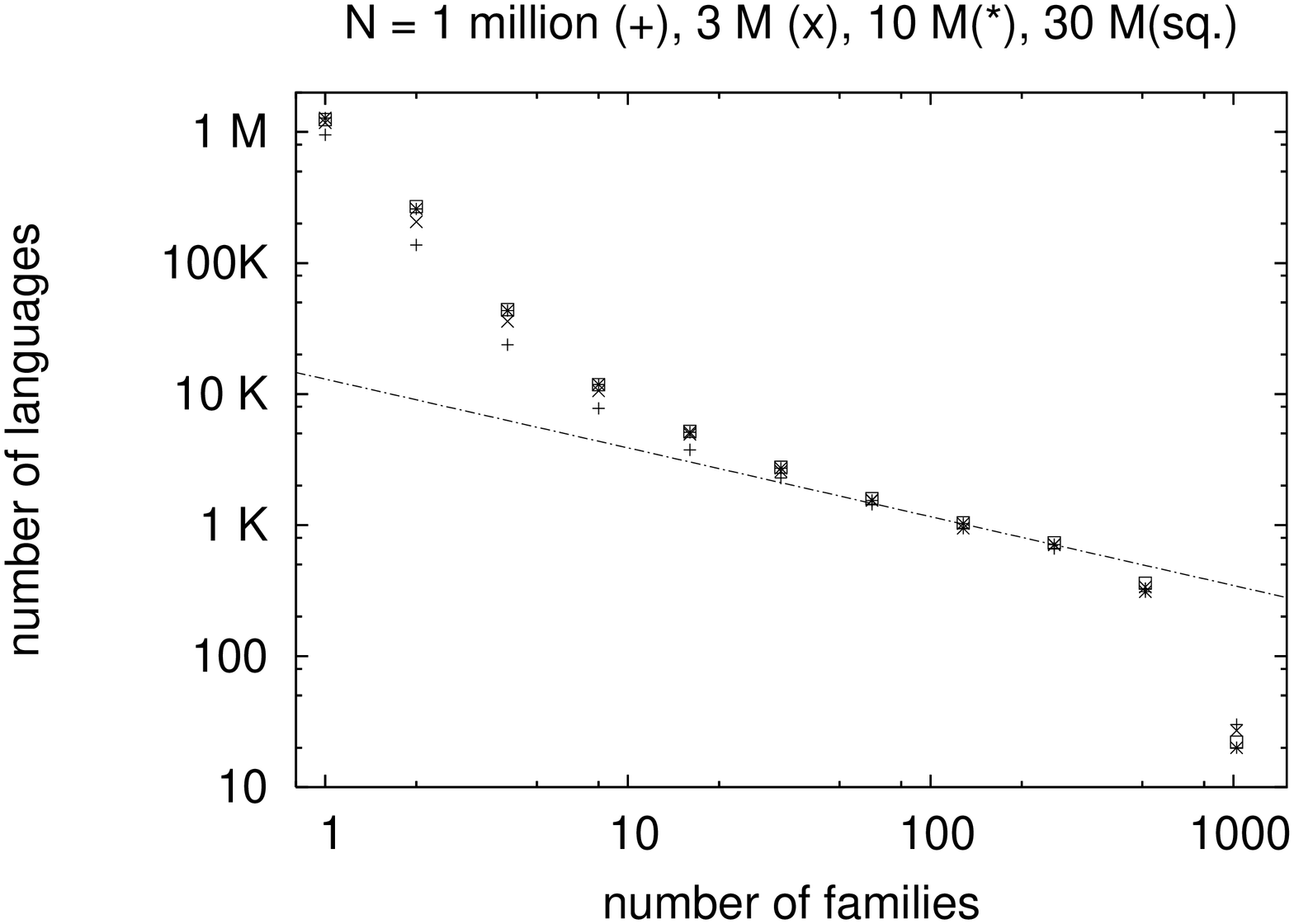}
\end{center}
\caption{
Number of families as a function of the number of different languages in this
family \cite{patt}. Top: Modified Viviane model for various lengths of the 
bit-strings. Bottom: Schulze model with $Q=5, F=8$ on directed Barab\'asi-Albert
scale-free network, $p=0.5, q=0.59, r=0.9$ for various population sizes. 
}
\end{figure}

\section{Other Models}

Years before physicists invaded en masse the field of linguistics, Nettle 
\cite{nettle} already wrote down a differential equation for the number
$L$ of languages,
$$dL/dt = 70/t - L/20  \quad ,$$
where time $t$ is measured in millennia. For long times, only one language 
(mathematically: zero languages) will remain; however that time lies far in the 
future. A more detailed splitting mechanism was introduced by Novotny and Drozd \cite{novotny} for the emergence of new languages from one mother language,
and gave a log-normal distribution of language sizes, in agreement with reality 
except presumably at the smallest sizes \cite{sutherland}. In the same spirit of
looking at languages as a whole, ignoring the individuals, are the very recent
models of Tuncay \cite{tuncay}, who coupled a splitting mechanism with random 
multiplicative noise in the size of the growing population, plus an extinction
probability, and found a power-law decay or a log-normal size distribution
for the simulated languages, depending on parameters. He also checked the 
lifetimes of the simulated languages. An ``early'' attempt to apply the
Ising model of statistical physics to linguistics \cite{canada} had 
little success.

Numerous coupled differential equations were studied by scientists coming from
theoretical chemistry, mathematics and computer science \cite{nowak} for the 
purpose of language learning by children. They have been applied \cite{newbook}
also to the competition of up to 8000 languages of adults, but since the
original authors have to our knowledge not followed this re-interpretation
of their learning model we now refer to \cite{nowak,newbook} for details and 
results. 

It was the population dynamics of Abrams and Strogatz \cite{abrams} which 
started the avalanche of physics papers on language competition. They 
assume two languages X and Y, spoken by the fractions $x$ and $y=1-x$ of a
fixed population with a time dependence 
$$ dx/dt = yx^as - xy^a(1-s) \quad ,$$
with a status or prestige variable $S$ which is close to one if X has a high
prestige and close to zero for low prestige of X. The neutral case is $s=1/2$.
The exponent $a=1.31$ was fitted to some empirical data of how minority languages
decay in size. If $a$ is replaced by unity we arrive at the logistic equation
of Verhulst from the 19th century, which was applied to languages by Shen
in 1997.

Fig.9 shows the resulting $x(t)$ if X is spoken initially by a minority of ten 
percent only. Then for low, neutral or slightly higher status $s$ of its 
language X, the fraction decays further towards zero, but for a higher status 
like $x = 0.7$ it finally wins over and is spoken by everybody (not shown). 
This may correspond to the influence of a colonial power; indeed in France today
most people speak French as a result of the Roman conquest of more than two 
millennia ago, and in Brazil many of the native languages have become extinct 
in the last five centuries since the Portuguese arrived there. 
%(In Germany the regions colonised by the Romans still speak German in order to 
%be understood by the barbaric tribes east of the Rhine river.)

\begin{figure}
\begin{center}
\includegraphics[angle=-90,scale=0.4]{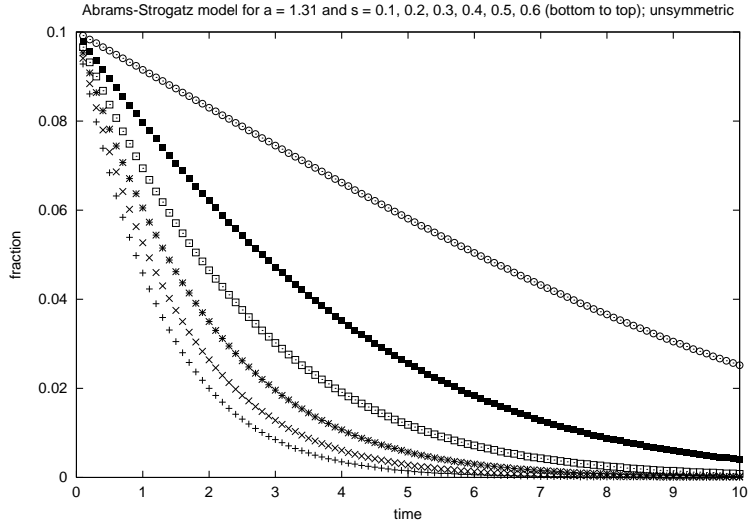}
\end{center}
\caption{Abrams-Strogatz model: Fraction of people speaking language X, versus
time, for an initial concentration of only ten percent and various prestige
values $s$ of this language. For $s \ge 0.7$ this language finally wins over
the other language Y. From \cite{newbook}. 
}
\end{figure}

This Abrams-Strogatz approach was soon generalized to a lattice by Patriarca
and Lep\"annen \cite{abrams}, and later to populations with bilingual speakers
\cite{spain,castello}, coexistence of the two languages \cite{argentina}
as well simulations based on individuals \cite{mallorca}.

Such agent-based simulations were also made by Kosmidis et al \cite{kosmidis}
who gave each person a string of 20 bits. The first 10 belonged to one language,
the last 10 to another language. In this way they were able to simulate 
people speaking, more of less correctly, one or two languages. One could also 
interpret their model as one for English which is a mixture of German
(Anglo-Saxon) and French words, due to the conquest of England by the Normans in
1066. 

Finally, Schw\"ammle \cite{schwammle} also used bit-strings, but to describe
biological ageing through the Penna model. The child can learn the languasge 
from the mother, the father, or both, thus also allowing for bilinguals. This
model allowed to simulate that languages are learned easier in youth than at
old age. In this way it builds a bridge between language competition and
language learning \cite{nowak}.

\section{How physics may inform linguistics: \\ prospects for future research}

As the research described above has progressed a larger design has become apparent, which consists in an empirical side looking for quantitative distributions involving languages \cite{wichmann,holman,wichhol} and a development of models simulating similar quantitive distributions. The hope is that as more and more quantifiable relations in and among languages are discovered and simulation models are developed which can adequately replicate these distributions, the simulation models will of necessity become more and more adequate as models of actual languages, and could therefore be employed for purposes beyond the ones for which they were designed. For instance, the revised Viviane model, which was designed to capture the distribution of speaker populations and the population of languages within families \cite{patt}, could potentially be employed for investigating absolute rates of language change, an issue with which linguists are very much concerned \cite{nichols}, inasmuch as knowledge of how fast languages change could provide us with a way to date prehistoric events involving people speaking given reconstructed languages. Thus, a strand of research where linguists and physicists can and will continue to cooperate is the search for quantifiable distributions on the one hand and the fine-tuning of models which can adequately simulate an increasing range of such distributions.

Apart from some exceptions \cite{spain,kosmidis,castello}, most work on language competition has assumed monolingual speakers. Since most of the world's population is bi- or multilingual this is clearly not adequate. Language shift will normally involve transitional bilingualism, or bilingualism may persist for centuries without the majority language necessarily replacing minority languages. Diglossia, i.e. the use of different languages for different purposes may help sustain bilingualism. Current models can be extended to investigate under which conditions bilingualism may persist or get reduced to monolingualism. Different kinds of situations can be modelled, such as the replacement of certain, but not all, languages within the domain of the Roman empire, the development of so-called linguistic areas, where several languages share a number of features (e.g., the Balkans, India, Mesoamerica), multilingualism caused by linguistic exogamy (the northwest Amazon region), the shift from one to another lingua franca with retention of minority languages (Mayan immigrants in urban United States shifting from Spanish to English but retaining Mayan languages), etc., and may be applied to situations where prehistoric interaction has left linguistic traces but where the nature of the interaction is unknown (e.g., the sharing of linguistic features around the entire coast of the Pacific Ocean). In the Appendix we provide a preview of the extension of the Schulze model to bilingualism.

An area where physicists may wish to try out their hands more is that of language change. Simulations may help linguists come to terms with realities that are accessible through empirical research only in small fragments. Languages develop and change through the interaction of multitudes of agents using large lexical inventories and complex grammars. The kinds of regularities that linguist can identify, such as the regularity of sound changes or directed paths of grammaticalization (roughly, the process whereby separate words become part of the morphology), are mostly accessible only to a retrospective view, trough the comparison of language stages dozens or hundreds of years apart. What lies between is a flux whose behavior is not easy to understand. 19th century historical linguists, with their focus on regular sound changes, lived in a universe of clean equations such as Latin $p$ = English $f$ (as in {\it pater = father}). The advent of 20th century sociolinguistics, with its focus on the social mechanisms behind sound changes, complicated the picture, much like the picture gets complicated when one moves from clean Newtonian physics to modern statistical physics, which tries to model the actual behavior and interaction of entities. Nevertheless, unlike physicists, linguists working on the way that languages change have taken little recourse to simulations that might help them understand the complexity of how language change or shift percolates within a community. For instance, a leading sociolinguist has argued that "networks constituted chiefly of strong ties function as a mechanism to support minority languages, resisting institutional pressures to language shift, but when these networks weaken, language shift is likely to take place" (p. 558 in \cite{milroy}). This hypothesis is based on just a few case studies, and such case studies are, one the one hand, extremely costly and, on the other, cannot even begin to cover the multitute of different situations that actually obtains. In addition to the strength of network ties, other important parameters are presumably the size of the group speaking the minority language, geography, prestige of one as opposed to the other language, economic gain involved in shifting language, age- and gender-determined mobility, and maybe more. The behaviors of such parameters can be investigated in simulations (e.g., for geography see \cite{reviews,quebec}).

Finally, more works needs to be done towards the integration of the modelling of language competition by physicists reviewed here and the modelling of language evolution by computational linguists \cite{cangelosi,briscoe2,christiansen,deboer,niyogi}. While physicists have been adept in modelling the interaction among agents but have operated with languages represented only by numbers or bit-strings, computational linguists offer elaborate grammar models. With more complex models of the interior structure of languages carried by agents, research need not be limited to a focus on language competition, but could be extended to issues of language structure itself. 

\section{Appendix: Bilingualism}

Several authors studied the possibility that people speak more than one 
language \cite{spain,kosmidis,castello}, and we do here the same for the
Schulze model on the square lattice, with $F=8$ features of $Q=3$ different
values, using only interactions to nearest neighbours \cite{barrier}. For
this purpose we modify the switch process.

Before, languages spoken by a fraction $x$ of the four neighbours were dropped
in favour of the language of a randomly selected neighbour with probability 
$(1-x)^2r \; (r=0.9)$. Now we do this at lattice site $i$ only if none of the 
four neighbours of $i$ speaks the mother language of site $i$: $x=0$; then
with probability $r$ we replace the mother language of $i$ by the mother 
language of a randomly selected neighbour. Otherwise, for $x > 0$, with the 
above probability $(1-x)^2r$, site $i$ learns as an additional ``foreign''
language a randomly selected (foreign or mother) language of a randomly 
selected neighbour. If in the latter case $x > 0$, site $i$ has already learned
a foreign language before, then this old foreign language is replaced by
the new foreign language.

\begin{figure}
\begin{center}
\includegraphics[angle=-90,scale=0.4]{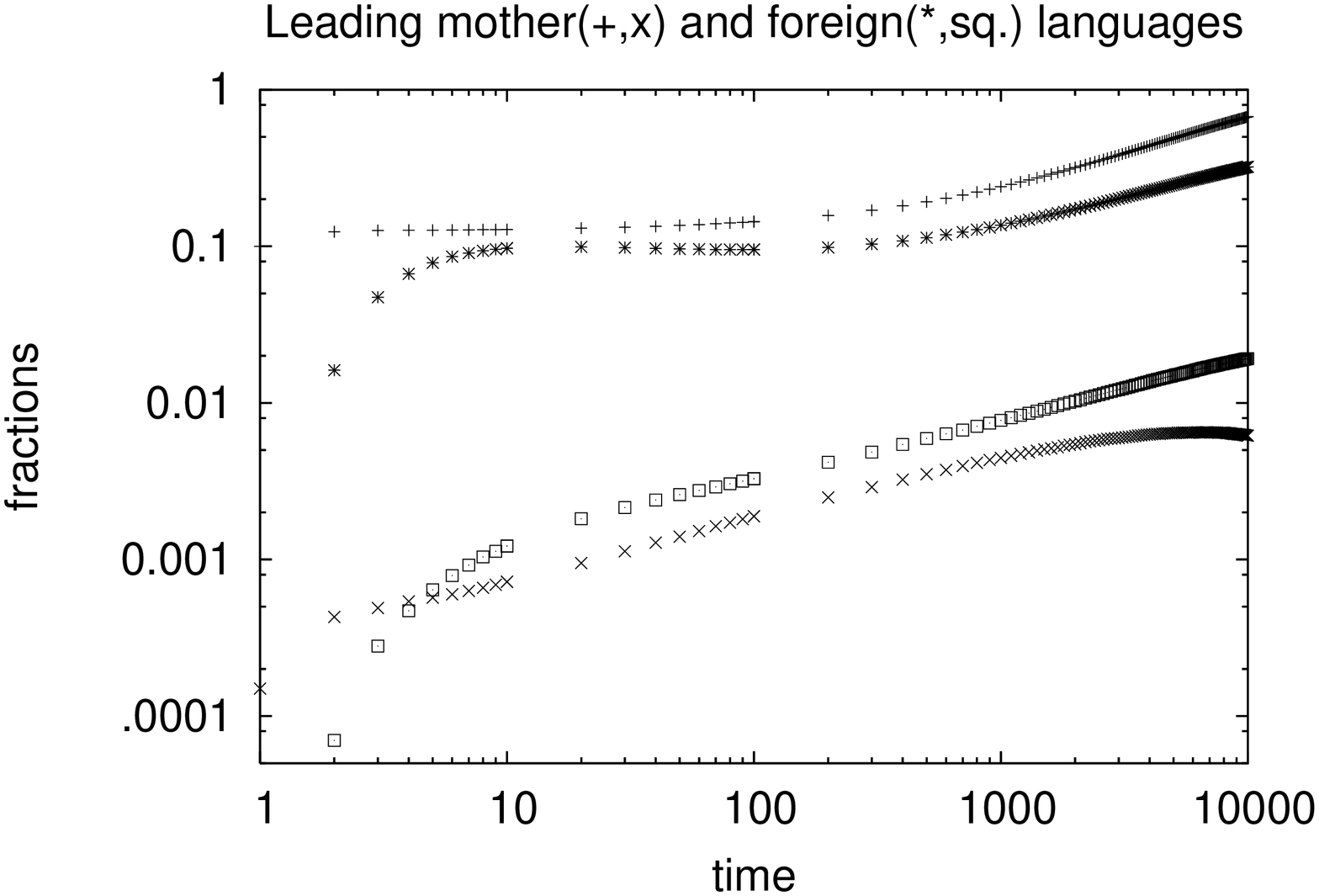}
\includegraphics[angle=-90,scale=0.4]{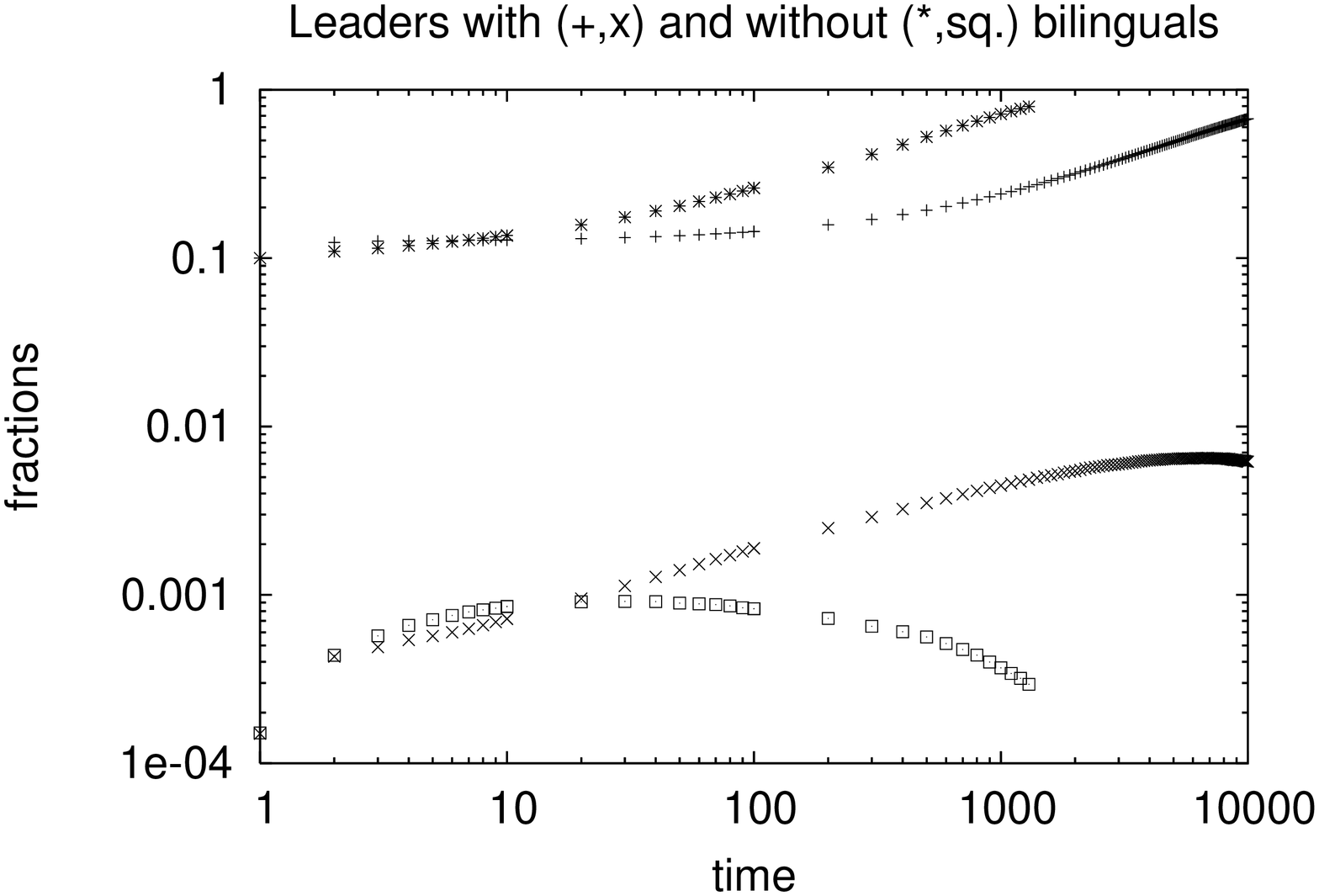}
\end{center}
\caption{Largest and second-largest languages in the Schulze lattice at $p=0.01,
 \; q = r = 0.9$ in a $8000 \times 8000$ lattice without migration. Part a
includes bilinguals, part b compares only the mother languages with and without
bilinguals.
}
\end{figure}
%langbil34.d, langbil00.d

\begin{figure}
\begin{center}
\includegraphics[angle=-90,scale=0.4]{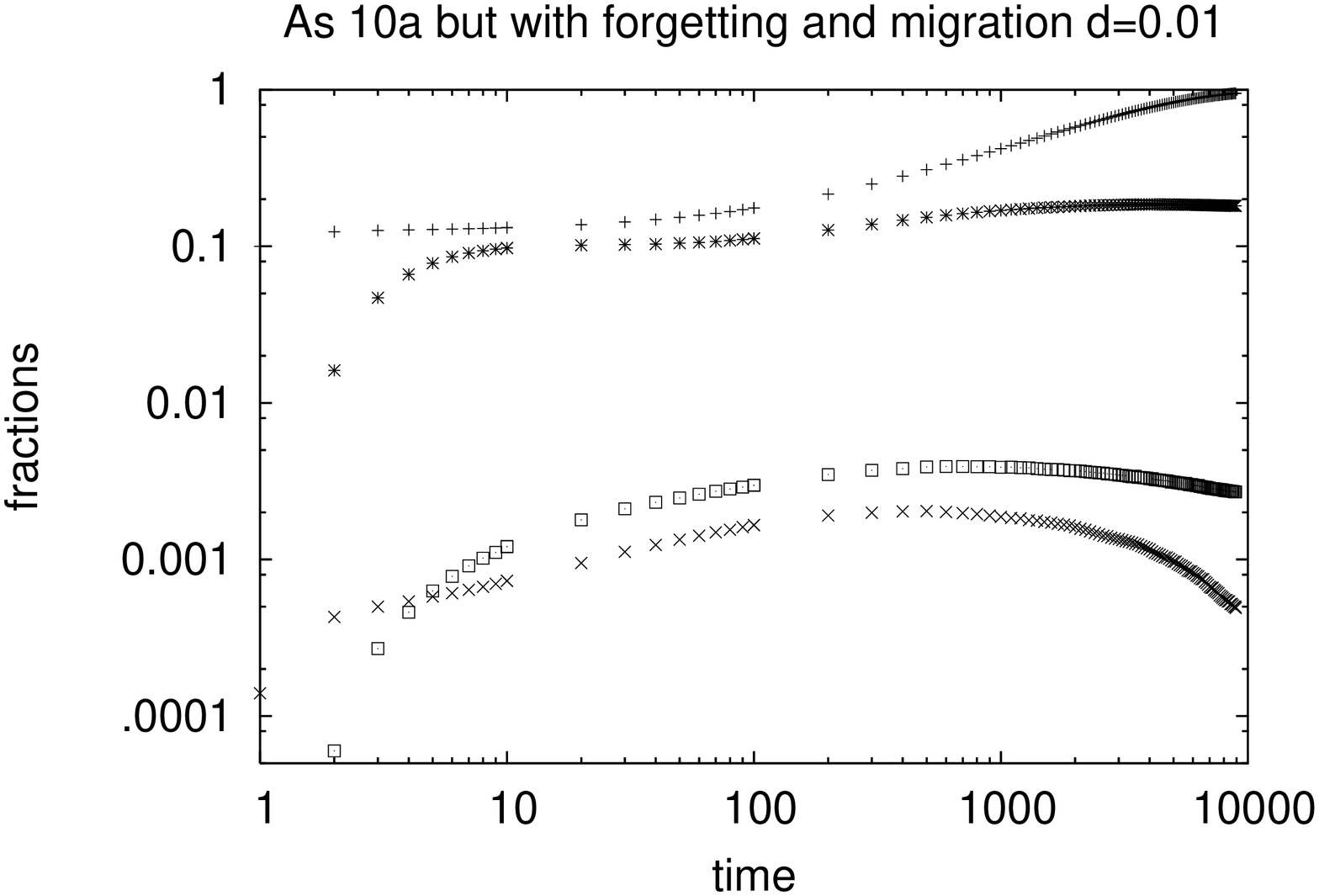}
\includegraphics[angle=-90,scale=0.4]{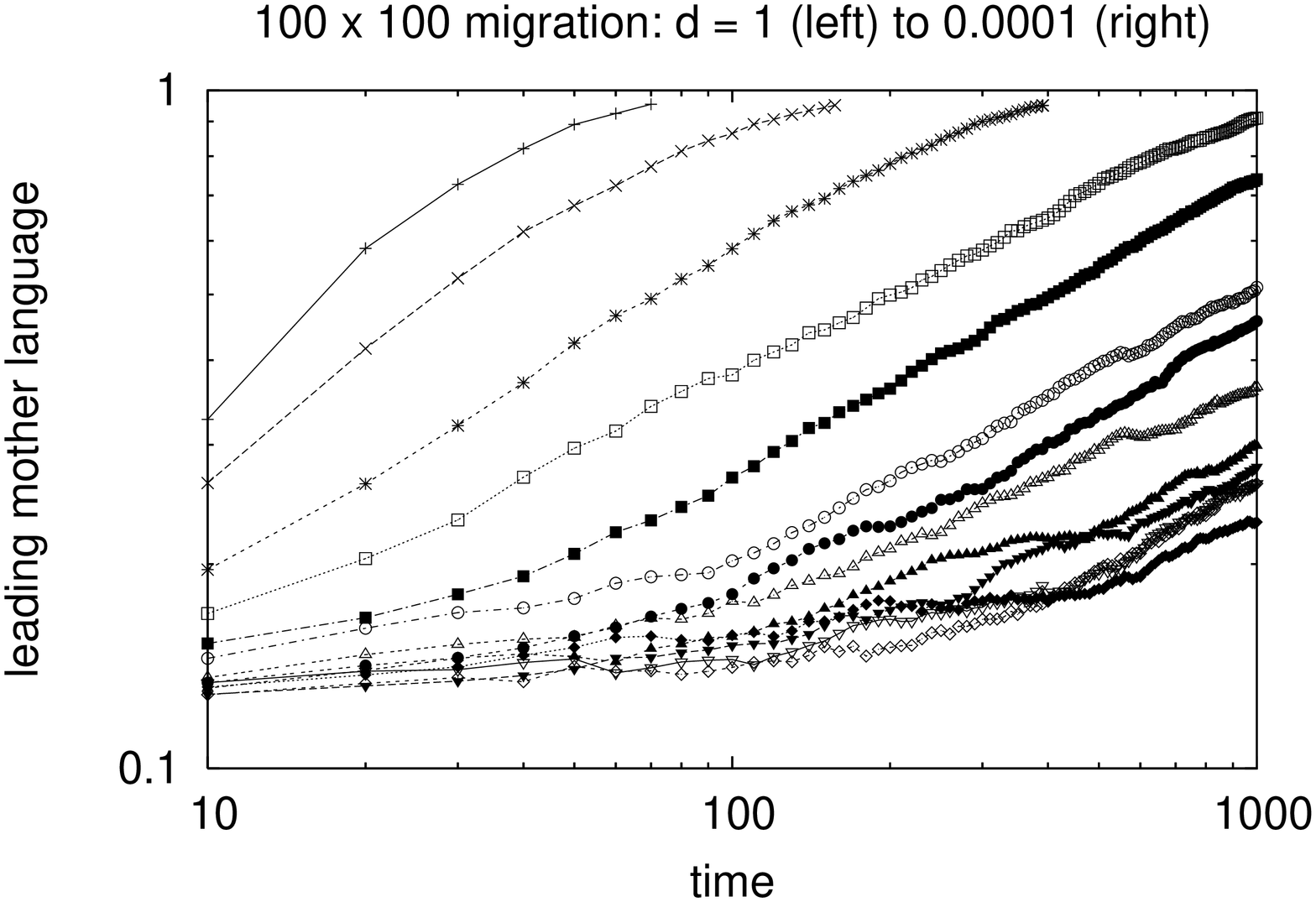}
\end{center}
\caption{Part a: As Fig 10a, but including a migration probability $d$ of 1 \%
and a site-dependent forgetting probability between zero and 5 \%,
on a $6000 \times 6000$ lattice. Part $b$ shows the drastic speed-up of 
dominance if the migration probability $d$ is enhanced: $d = 1, 0.5, 0.2, 0.1,
\dots, 0.0005, 0.0002, 0.0001$.
}
\end{figure}
%langbil60.d, langbil64*.d

These are the learning and replacement events if everybody can speak at most two
languages. If instead the number of languages for each person is restricted
by an overall upper limit, then for $x > 0$ the last-learned foreign language is
replaced by the newly selected neighbour language. If this upper limit is set
equal to one, we go back to the model of monolingual speakers. In all cases,
$x$ is the fraction of neighbours of $i$ speaking as their mother tongue the 
mother language
of site $i$. For language diffusion we took $q=0.9$ throughout, and for language
change mostly $p=0.01$. Thus one iteration may correspond to about one human
generation. 

We start with a fragmented distribution of mother languages and no foreign
languageas, except that one particular language is spoken initially by ten 
percent of the people, randomly spread over the lattice. Then we check if this
``lingua franca'' finally (after at most $10^5$ iterations) is spoken by
about everybody: transition from fragmentation to dominance of initially 
favoured language. (If another language dominated we count this case as
fragmentation.) 

For ten $50 \times 50$ lattices, this transition happened up to $p=0.04$ if
bilinguality is allowed, while for monolinguals it happens up to the larger
changing rate $p = 0.09$: Bilinguality makes dominance of one language less
stable against continuous changes; see also \cite{castello}. For $Q=5$ instead 
of 3 this limit shifts from 0.04 to 0.05, while for $Q = 3, \; F = 16$ it is 
about 0.03.  Fig.10 shows for $8000 \times 8000$ lattices the time dependence 
of the fraction of people speaking the largest and the second-largest language, 
separately for mother tongue and foreign languages; the comparison for the case
without bilinguals is restricted to the mother language and shows that dominance
is reached faster without bilinguals. 

In these simulations, after a short time everybody speaks two languages, and if 
up to ten languages are allowed, then again after a short time everybody speaks
ten languages. This is nice but unrealistic. In order to take into account that 
people forget again foreign languages which were learned but not used, or give
up learning a foreign language when the need for it dissipates, we assume that
at every time step each speaker (more precisely, each lattice site) may give up 
the last-learned foreign language if none of the neighbours at that time speaks
this language. This forgetting happens with a probability between zero and five 
percent, fixed for each site randomly and independently at the beginning. 

In addition we included migration via exchanging locations: A speaker
or family exchanges residence with a randomly selected neighbour, and both
carry their languages with them. This happens at each iteration with a
probability $d$; physicists call $d$ the diffusion constant. Fig.11 shows that
appreciable migration can drastically speed up the growth of the lingua
franca from having an initial advantage of being spoken by ten percent of the 
population to being dominant.

\end{document}